# Superior dielectric breakdown strength of graphene and carbon nanotube infused nano–oils

**Purbarun Dhar, Ajay Katiyar[*], Lakshmi Sirisha Maganti, Arvind Pattamatta and Sarit K Das[#]**

Department of Mechanical Engineering, Indian Institute of Technology Madras,
Chennai – 600036, India

*Research and Innovation Centre (DRDO), IIT Madras Research Park,
Chennai – 600113, India (2[nd] affiliation of author)

#School of Mechanical, Materials and Energy Engineering (SMMEE),
Indian Institute of Technology Ropar, Rupnagar–140001, India (Present address of author)

## ABSTRACT

**Nano–oils comprising stable and dilute dispersions of synthesized Graphene (Gr) nanoflakes and carbon nanotubes (CNT) have been experimentally observed for the first time to exhibit augmented dielectric breakdown strengths compared to the base transformer oils. Variant nano–oils comprising different Gr and CNT samples suspended in two different grades of transformer oils have yielded consistent and high degrees of enhancement in the breakdown strength. The apparent counter–intuitive phenomenon of enhancing insulating caliber of fluids utilizing nanostructures of high electronic conductance has been shown to be physically consistent thorough theoretical analysis. The crux mechanism has been pin pointed as efficient charge scavenging leading to hampered streamer growth and development, thereby delaying probability of complete ionization. The mathematical analysis presented provides a comprehensive picture of the mechanisms and physics of the electrohydrodynamics involved in the phenomena of enhanced breakdown strengths. Furthermore, the analysis is able to physically explain the various breakdown characteristics observed as functions of system parameters, viz. nanostructure type, size distribution, relative permittivity, base fluid dielectric properties, nanomaterial concentration and nano–oil temperature. The mathematical analyses have been extended to propose a physically and dimensionally consistent analytical model to predict the enhanced breakdown strengths of such nano–oils from involved constituent material properties and characteristics. The model has been observed to accurately predict the augmented insulating property, thereby rendering it as an extremely useful tool for efficient design and prediction of breakdown characteristics of nanostructure infused insulating fluids. The present study, involving experimental investigations backed by theoretical analyses and models for an important dielectric phenomenon such as electrical breakdown can find utility in design of safer and more efficient high operating voltage electrical drives, transformers and machines.**

## 1. INTRODUCTION

Carbon based meso–nanomaterials; viz. graphene (Gr), carbon nanotubes (CNT), fullerenes, nano–diamonds etc. have shown tremendous potential to revolutionize the modern technological prowess of mankind. Gr and CNTs have been utilized to develop advanced systems along avenues of smart materials [1-3], sensors [4-6], MEMS/NEMS [7, 8], modern nanoelectronics [9-11], power systems [12, 13] and many more. In recent times, variant phenomena, smart applications and utilities of such carbon nanostructures in stable, colloidal phases have also been probed [14-17]. However, a majority of such endeavors have been limited to understanding the chemistry of such nanostructures in the colloidal state or their ability to tune some special properties of the fluid phase, which, majorly has been concentrated on the thermophysical properties of the fluid [18-20]. The fact that nanomaterials such as Gr and CNT possess exceedingly high electron transport characteristics in a manner points to the fact that their presence in a colloidal system could be engineered to tune or modify the material dielectric properties of the fluid phase.

Incorporation of inorganic/ organic additives to fluids in order to modify their dielectric characteristics has been practiced for a reasonably long time. Among the dielectric characteristics, researchers have often focused their attention towards modifying the dielectric breakdown strength of insulating fluids, since it is one of the most important aspects that govern safe operation of modern power systems which are cooled by insulating oils, ranging from transformers, large scale super capacitors, resonator coils, high charge density components etc. However, chemical additives often suffer from inherent disadvantages, such as corrosion of components, possible reaction at elevated electric field intensities and temperatures, possibility of flammability in case of corona discharge at high voltages etc. Consequently, in recent times, researchers have started looking for alternatives to such additives in the form of nanomaterials [21, 22]. Nanomaterials offer certain advantages over the conventional additives, viz. high stability as dispersions, negligible corrosive action, expanded options over material selection and high dielectric and electronic properties which are of utmost importance for modifying the dielectric characteristics of the base fluid.

Among the few reports available in the domain of enhancing the dielectric breakdown (DB) strength employing stable dispersions of nanostructures in insulating fluids, the nature of nanomaterial utilized can be divided in to two broad categories, semiconducting (titanium dioxide [23-26]) and super paramagnetic (iron (II,III) oxide [27-29]). Both the categories of particles have been resorted to since such materials possess high values of relative permittivity, which in turn leads to rearrangement of the electric field within the fluid and electron scavenging in a manner so as to reduce the possibility of DB at the expected field intensity. Furthermore, all such reports are completely experimental in nature and very limited insight into the actual physics involved in the augmented transport property can be extracted. Purely theoretical analysis [30] reveals that augmentation of breakdown voltage by scavenging of free electrons is also possible with nanostructures of exceedingly high electrical conductivities. Hence, intuitively, Gr and CNT based insulating fluid dispersions are expected to exhibit high enhancements in the DB voltages. The present paper investigates experimentally the augmented DB strength of Gr and CNT based transformer nano–oils, as a function of nanostructure size, concentration, electronic and dielectric properties and nano–oil temperature. Thorough survey of existing literature on the subject matter reveals that the present article is the first of its kind to report carbon based nanomaterials as potentials for enhancing breakdown strengths of insulating fluids. Based on experimental observations and mathematical analysis of the electro–hydrodynamics involved, a comprehensive picture of the underlying physics and mechanisms has been put forward. Moreover, given the high probability of direct applications of such nano–oils in high charge density/ high voltage electrical machines and components, an analytical model has been proposed to predict the augmentations possible for a particular nano–oil or efficient design of such nano–oils as per necessity. The physically and dimensionally consistent model derived from first principles accurately predicts the enhanced breakdown voltages, as validated against experimental data.

## 2. MATERIALS AND METHODOLOGIES

### 2.1. NANOMATERIALS: SYNTHESIS AND CHARACTERIZATION

The nanomaterials utilized in the present study comprises two samples of nano–graphene synthesized in–house and two samples of multi–walled carbon nanotubes (MWCNT) procured from firms manufacturing and supplying nanomaterials. All involved in–house synthesis has been carried out with commercially available reactants and reagents which have been used as obtained and without further purification. The graphenes (Gr) utilized have been synthesized chemically in the form of reduced graphene oxide (rGO), which has been obtained from graphene oxide (GO) synthesized via the modified Hummer's method [19]. In a typical synthesis procedure, 1 gm of graphite (G) powder (average flake size ~ 20–50 μm) is initially added to ~ 25 mL conc. sulphuric acid (98%) and allowed to stand at 90 oC for 2 hours. Next, 2 gms each of anhydrous potassium peroxosulphate and anhydrous phosphorus pentoxide is added to the mixture and the same stirred vigorously for 3 hours while maintaining the initial temperature for pre–oxidation of the graphite flakes. The resulting mixture is carefully centrifuged to obtain the preoxidized graphite (pG) as sediment. The same is washed thoroughly with deionized (DI) water and further centrifuged to collect the pG, which is then dried overnight in a hot air oven. The next step, comprising oxidation of the pG to GO involves dissolution of the pG in ~ 50 mL conc. sulphuric acid (98%) and the mixture stirred vigorously while maintaining ice cold conditions by employing an ice bath. The ensuing process is highly exothermic and an ice cold atmosphere is compulsory in order to avoid explosive reaction. To the mixture, 3 gm of potassium permanganate is added in small batches over a period of time. The reaction yields an algae–green mixture and the vigorous stirring under ice cold conditions is continued for 6–7 hrs. Further, the mixture is diluted by adding ~ 500 mL DI water slowly and stirred till the mixture attains room temperature. The oxidation process is finalized by adding 10 mL hydrogen peroxide (5%) to the solution. The formation of a bright yellow solution confirms the formation of GO. The solution is allowed to stand overnight for the GO to sediment. The precipitate is collected via decantation and washed thoroughly with DI water. The resulting mixture is centrifuged at ~ 12000 rpm to collect the GO, which is then added to pure acetone (concentration ~ 0.1 wt. %) and ultrasonicated to form stable suspensions. The suspensions are then spread uniformly as thin films onto petri dishes and oven–dried overnight. The dried films are scraped off with scalpels to obtain pure, non–aggregated GO.

The dried GO is dispersed in DI water (0.1 wt. %) and ultrasonicated to form stable suspensions, which are then heated to 60–70 °C. To 250 mL of the solution, either 250 mg of finely powdered sodium borohydride or 25 mL of hydrazine hydrate and 10 mL ammonia solution is added, while stirring for 1 hr. at the aforementioned temperature. Either of the mentioned processes can be utilized in order to reduce the functional groups on the GO to obtain rGO/Gr suspension. The suspensions are centrifuged at ~ 12000 rpm to obtain the Gr, which is further washed, centrifuged and suspended in pure acetone and dried to obtain non–aggregated Gr nano–flakes similar to that mentioned in the preceding paragraph. The synthesized Gr has been subjected to in–depth characterization so as to comprehend the physical parameters. The detailed characterizations for the Gr samples have been illustrated in Fig. 1.

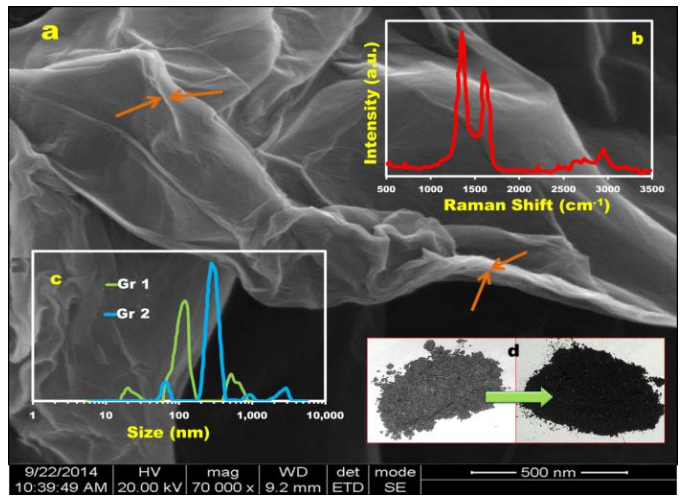

**Figure 1:** Characterization of the graphene samples, namely Gr1 and Gr2. (a) Illustrates the High Resolution Scanning Electron Microscope (HRSEM) image of the synthesized Gr. The fabric mimicking texture and the nanoscale folding and wrinkles on the sheets, as indicated by arrows, qualitatively indicates exfoliation of G to form Gr. (b) Raman spectroscopy intensity vs. shift plot for Gr. The presence of the unique D, G and 2D bands at ~1350 cm-1, 1600 cm-1 and 2800 cm-1 respectively within the spectrum establishes the presence of Gr. (c) Dynamic Light Scattering (DLS) analysis for the samples reveals the polydisperse size distribution for the samples, wherein, the micron scale G flakes has been completely converted to nanoscale Gr, with the exception of a small population among Gr2 in the 2–4 micron range. Gr1 and Gr2 exhibit a maximum flake population at ~ 100 nm and 300 nm respectively. (d) Visual appearance of the synthesized Gr (right, black) from the natural G (left, grey).

The synthesized Gr has been characterized qualitatively from its visual appearance and SEM image and quantitatively from analysis of the Raman spectra and DLS data. The SEM image reveals presence of nanoscale thin flakes, highly convoluted and mimicking folds on sheets of fabric, which indicates the presence of graphenic structures. G being multilayered, do not exhibit such pristine convolutions and folds. The oxidation followed by reduction exfoliates the layers, simultaneously creating folds, creases and wrinkles on the individual flakes. Hence, presence of such nanoscale folds and wrinkles (represented in Fig. 1(a) by arrows) qualitatively confirms transformation of graphitic structures to Gr. Quantitatively; the Raman spectrum is an effective tool to confirm the presence of Gr. As illustrated in Fig. 1(b), the spectrum exhibits the characteristic signature peaks of graphene systems. The D (~1350 cm-1) and G (~1600 cm-1) bands represent planar defects in the sheets and the in–plane stretching of the sp2 hybridized carbon atoms in Gr respectively. Further, the existence of the 2D band (~2800 cm-1) within the spectrum confirms the presence of Gr. The ratio of intensity of the 2D band to that of the G band for the present samples indicates Gr systems of less than equal to five layers [31], i.e. the dielectric and electrical transport phenomena unique to Gr are expected to manifest themselves during the present experimentations. Two distinct Gr samples, viz. Gr1 and Gr2 have been synthesized. The samples are distinguished based on their polydispersity, which has been characterized utilizing DLS (as illustrated in Fig. 1(c)). Analysis reveals that Gr2 has a much larger average flake size than that of Gr1.

The MWCNT systems utilized in the present study have been procured from Nanoshel Inc. (USA) (manufactured via Chemical Vapor Deposition (99.5 % purity), illustrated in Fig. 1 (b)) and the characteristic dimensions of the nanostructures have been verified utilizing Transmission Electron Microscopy (TEM). The TEM images of the CNT samples have been illustrated in Fig. 2. The nanotube outer diameters for the two samples utilized, CNT1 and CNT2, have been verified against the manufacturer's claims and illustrated in Fig. 2(c) and (d). The average diameter for CNT1 and CNT2 are 20 nm and 10 nm respectively whereas that specified by the manufacturer was 20–40 nm and 5–20 nm respectively, thereby confirming the manufacturer's claims. The aspect ratio lies within the range 100–400 and 200–500 for CNT1 and CNT2 respectively as per the characterization report provided by the manufacturer. The nanomaterials synthesized and/or procured have been utilized to prepare nano–oils utilizing two types of transformer oils procured from certified dealers. Two grades of transformer oils have been utilized and have been denoted as Oil1 and Oil2 hereafter. The nano–oils have been synthesized by adding the required amount of nanomaterial to the oil (in wt. %) followed by ultra-sonication for 1 hour. Oleic acid (OA) has been utilized as the capping agent/ surfactant for samples with nanomaterial concentrations of and above 0.05 wt. % so as to induce more stability to the suspensions. The amount of OA utilized in each case varies from 0.5-2 mL per 500 mL of oil depending upon the concentration of nanomaterials. It is noteworthy that the presence of OA in the oil in such dilute proportions has negligible effect on the breakdown strength of the base oil; as established from experimental observations. Also, the average shelf life of the dilute samples ranges above 1 month when kept undisturbed. Samples have been denoted hereafter in accordance to its constituents, e.g. Gr2 and Oil1 based nano–oil has been named Gr2 Oil1. Illustrations of the base transformer oil, CNT and Gr based nano–oils have been provided in Fig. 3 (c), (d) and (e) respectively.

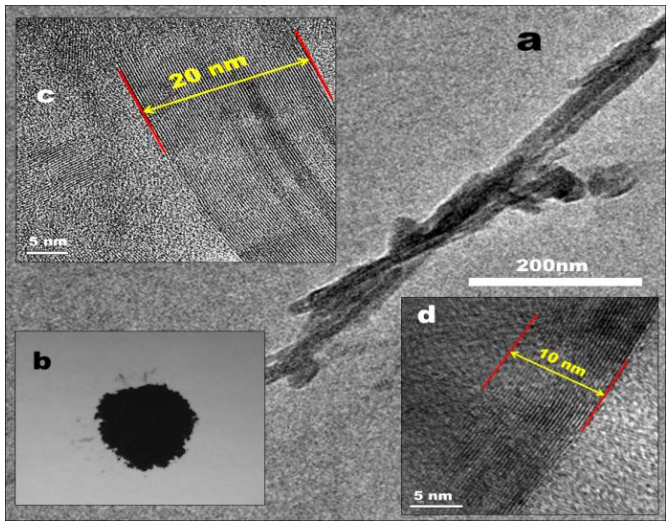

**Figure 2:** Characterization of the MWCNT samples, namely CNT1 and CNT2. (a) TEM image of the MWCNT samples confirming tube structure (b) Image of the procured sample (c) TEM image of CNT1 (20 nm average outer diameter) (d) TEM image of CNT2 (10 nm average outer diameter).employment history and special fields of interest.

## 2.2. INSTRUMENTATION

Determining the DB strength of the transformer oils and the nano–oils requires a voltage measurement instrument operated in accordance to specified ASTM/IEC standards and codes. For the present study, a custom manufactured automated high voltage (HV) test compartment has been utilized and components of the same have been illustrated in Fig. 3 (a). The setup consists of two compartments, viz. sample basin holding compartment with HV transformer and terminals and the controller circuit and actuator compartment. The equipment utilizes single phase 220 V AC as the input and a HV transformer steps up the voltage as governed by the automated controller. The HV is fed across the electrode terminals with the test fluid in between and a precision micro–ammeter reads the current passing across the terminals. At the moment of DB, an arc of high current is discharged across the electrodes and the controller circuit trips the HV circuit. The voltmeter reading at that tripping point is read as the breakdown strength of the dielectric fluid. The tests have been performed in accordance to the ASTM D–877 utilizing two hemispherical electrodes of 5 mm diameter (the apex points of the hemispheres face each other) with electrode spacing of 5 mm (the spacing is fixed utilizing specific gauges). The HV across the terminals is raised by an automated circuitry at an approx. rate of 2 kVs-1. The DB strength of each set of oil is determined 10 times and the arithmetic mean is taken as the breakdown strength of the same. Experiments have also been performed to deduce the effect of temperature on the DB strength of nano–oils, wherein, the oil is first heated to elevated temperatures and the experiments performed thereafter. The temperature of the oil in between the gaps of the electrodes has been monitored utilizing a precision thermometer and the temperature difference from the required value has not been allowed to traverse beyond ± 2 oC. The mean of 10 readings at and around (within the allowed tolerance) a particular temperature has been considered as the DB voltage at that temperature.

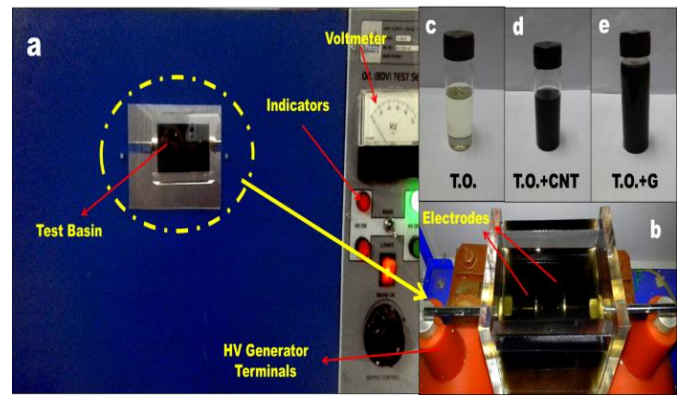

**Figure 3:** Components of the dielectric breakdown setup (a) Front view of the test setup. The indicators exhibit the status of the high voltage (HV) terminals across the sample (red lamp indicated active HV circuit whereas green indicates dormancy), which can be viewed through the safety window. The system is automated to increase or decrease the voltage across the terminals (displayed by the HV voltmeter; rated 100 kV with least count of 200 V) at the desired rate. (b) The sample holding basin with the nano–oil and the electrodes assembled. The protruding ends of the electrodes are rested on the HV terminals. (c) Transformer oil (T.O.) (d) CNT based nano–oil (e) Graphene based nano–oil.

## 3. RESULTS AND DISCUSSIONS

### 3.1. Augmented dielectric breakdown strength

Experimental observations and analysis reveals that the presence of Gr and CNT in transformer oils, when in the form of a stable suspension with appreciably long shelf life, leads to augmented DB strength of the oil when compared to its virgin form. Oil1 and Oil2 have been observed to possess DB strength of ~ 30 kV and 40 kV respectively when tested with the 5 mm electrode gap configuration (i.e. 6 kV/mm and 8 kV/mm strength respectively). However, since the electrode gap has been maintained constant for all the present experiments, the BD strengths have been expressed in their potential form rather than the field intensity form. The BD characteristics of Gr1 and Gr2 based nano–oils has been illustrated in Fig. 4(a) and (b) respectively and the respective enhancements and critical concentration at which maximum enhancement is obtained has been illustrated in Fig. 5. As evident, the presence of Gr nanoflakes in the oil leads to enhanced insulting properties, with a maximum augmentation of ~ 45% observed for Gr2 Oil2 nano–oil at very dilute dispersion concentration of 0.075 wt. %. Likewise, Fig 6(a) and (b) illustrate the breakdown performance of CNT based nano–oils with the respective enhancements and concentration corresponding to peak position illustrated in Fig. 7. In case of CNTs, the observed maxima in enhancement of breakdown voltage exceeds the capability of Gr, with ~ 50% enhancement observed for CNT2 Oil1 nano–oil at a more dilute concentration of 0.025 wt. %.

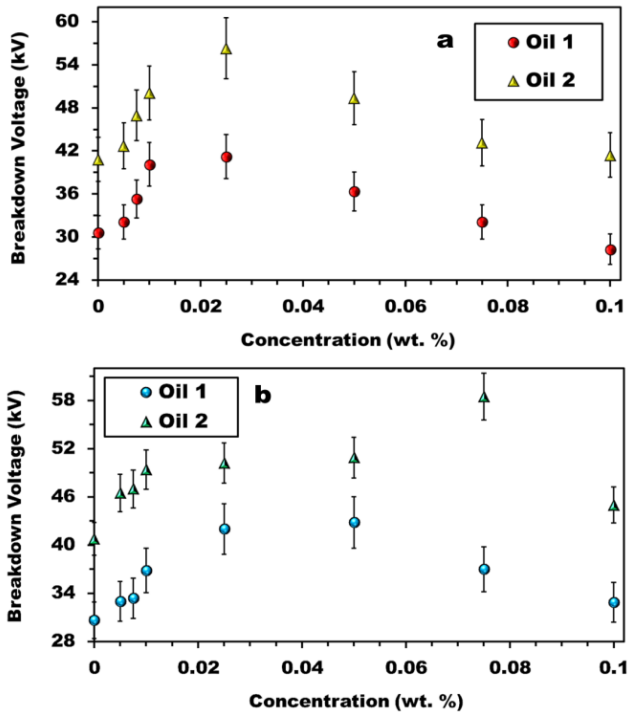

**Figure 4:** Dielectric breakdown characteristics of graphene based nano–oils **(a)** Gr1 based nano–oils **(b)** Gr2 based nano–oils.

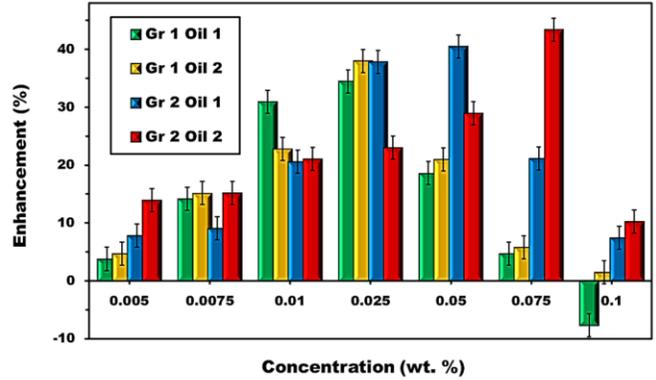

**Figure 5:** Magnitudes of augmentation of the DB strength of graphene based nano–oils, comparison between various samples and critical concentrations for maximum obtainable enhancements.

The phenomenon of enhanced DB strength of dielectric oils upon addition of nanostructures possessing very high electrical conductivities such as Gr and MWCNT, though counterintuitive at first glance, is physically consistent and can be explained qualitatively based on the electrodynamic processes within the nano–oil when exposed to HV stresses. The superior DB voltage exhibit by the nano–oils is a direct consequence of the morphed electrodynamics within the oil caused by the presence of the nanostructures, thereby making the electrical response of the nano–oil grossly different from that of the base oil. Among variant parameters, the extent of morphing of the dielectric response of the system can be deduced qualitatively from the charge relaxation parameter for the system. In brief, DB within a liquid occurs due to the establishment of a narrow, highly conductive 'tunnel/ channel' which propagates outwards from the electrodes, the merger of which leads to the formation of a highly ionized pathway for the charge to propagate across the dielectric medium [32, 33]. The transient development of the zones of ionized molecules that propagate towards each other (known as streamers) is governed by the DB strength of the liquid and the growth period is directly proportional to the DB strength and the applied electric field intensity. Thereby, if the growth and/or response period of the ionic constituents responsible for development of the streamers can be reduced or delayed, the probability of DB for the liquid at a given electric stress is reduced accordingly. This is the possible mechanism by which the presence of nanostructures enhances the DB strength of the oil. In the event that the charge relaxation time constant of the nanostructures is markedly lesser the characteristic time of streamer development and propagation, the free electrons released due to near–field ionization at the electrodes is effectively captured or scavenged by the nanostructures, thereby appreciably preventing the growth of streamers to the requisite strength. A qualitative illustration of the mechanism has been provided in Fig. 8 (a). Au contraire, the converse leads to negligible or even adverse effects on the DB strength of the liquid and the possible mechanism has been discussed in subsequent portions.

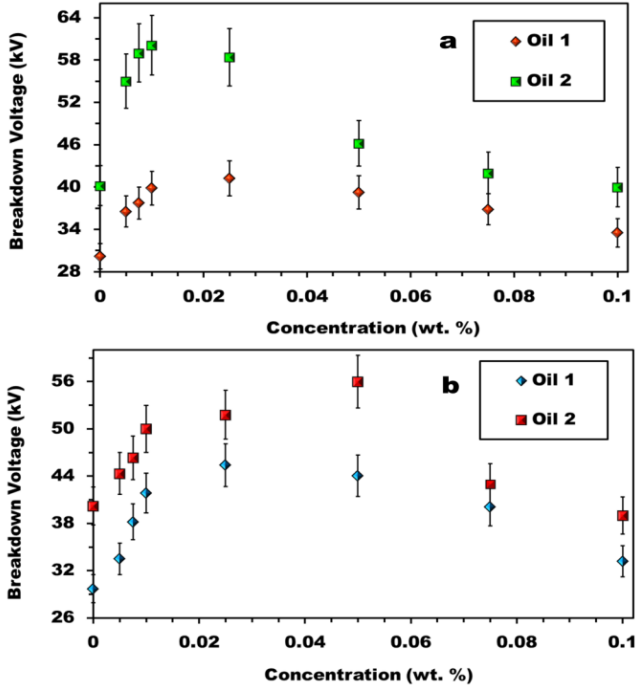

**Figure 6:** Dielectric breakdown characteristics of graphene based nano–oils **(a)** CNT1 based nano–oils **(b)** CNT2 based nano–oils.

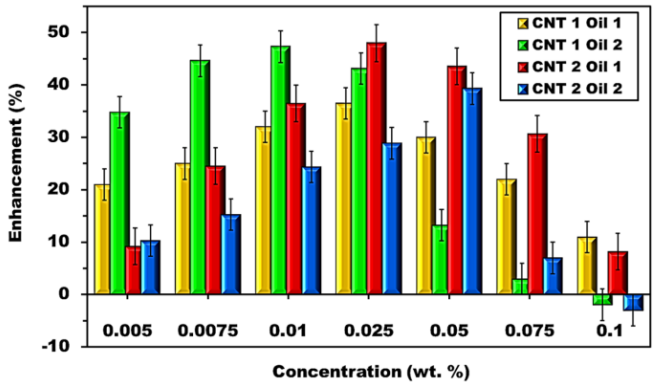

**Figure 7:** Magnitudes of augmentation of the DB strength of graphene based nano–oils, comparison between various samples and critical concentrations for maximum obtainable enhancements.

A qualitative assessment of the efficacy of a given nanostructure in augmenting the breakdown strength of a liquid can be made based on the charge relaxation time ($\tau$), absolute permittivity ($\varepsilon$) and electrical conductivity ($\sigma$) of the nanostructure (subscripted $np$) in conjunction with the surrounding fluid (subscripted $f$) [30]. Initially, a nanostructure (of characteristic length '$d_{np}$') situated at a spatial location describable by $(r,\theta,z)$ coordinates at a frozen instant of time ($t$) is considered, such that the origin of the describing polar coordinate system coincides with the geometrical center of the nanostructure. An electric field ($E_0$), such that the field lines are directed collinear to the z–coordinate axis, with the nano–oil as the participating medium, is applied between the electrodes. In absence of nanostructures, the field is expected to distribute uniformly within the oil and ionize the oil in the region between the electrodes homogeneously. However, the presence of stably suspended nanostructures leads to disruption of the uniformity of the field in the vicinity of the nanostructure–oil interface, leading to marked different electrical repose of the oil in between the electrodes. The electric field, when solved in accordance to the Laplace equation establishing null divergence of the field potential, yields the radial and polar components of the spatio–temporal distribution of the electric field within the nanostructure as [30]

$$E_r(r,\theta) = E_0 \cos\theta \left[ 1 + \left\{ \left(\frac{d_{np}^3}{4r^3}\right)\left(\frac{\varepsilon_{np}-\varepsilon_f}{\varepsilon_{np}+2\varepsilon_f}\right)e^{-t/\tau} \right\} + \left\{ \left(\frac{d_{np}^3}{4r^3}\right)\left(\frac{\sigma_{np}-\sigma_f}{\sigma_{np}+2\sigma_f}\right)(1-e^{-t/\tau}) \right\} \right]$$

(1)

$$E_\theta(r,\theta) = E_0 \sin\theta \left[ -1 + \left\{ \left(\frac{d_{np}^3}{8r^3}\right)\left(\frac{\varepsilon_{np}-\varepsilon_f}{\varepsilon_{np}+2\varepsilon_f}\right)e^{-t/\tau} \right\} + \left\{ \left(\frac{d_{np}^3}{8r^3}\right)\left(\frac{\sigma_{np}-\sigma_f}{\sigma_{np}+2\sigma_f}\right)(1-e^{-t/\tau}) \right\} \right]$$

(2)

In the above equations, the charge relaxation time of the nanostructure with respect to the fluidic phase is expressible as

$$\tau = \frac{2\varepsilon_1 + \varepsilon_2}{2\sigma_1 + \sigma_2}$$

(3)

Considering Gr based nano–oils, an estimate of the relaxation time can be performed. The relative permittivities of the oils have been determined utilizing a custom built parallel plate capacitor system and a multi–frequency LCR meter (HP, USA). The capacitor is first calibrated against air and a graded mineral oil with standard set of data available. Experiments reveal that at frequency ranges of 10–100 Hz (prioritized in accordance to the generation and distribution frequencies utilized worldwide) Oil1 and Oil2 possess relative permittivities of ~2.25 and ~2.48 respectively. The electrical conductivity of the oils has been considered to be of the order of $10^{-12}$ S/m, as obtained from published data [30]. In similitude, the relative permittivity of the Gr samples have been considered equivalent to that of graphitic systems (ranging from 2–2.5) and the electrical conductivity has been determined based on the room temperature electronic mobility in impure (to keep the predictions inclined towards the worst case scenario side) graphene sheets, which is of the order of $10^8$ S/m. Based on the above discussions, the magnitude of '$\tau$' for the Gr based nano–oils has been predicted to be ~ $10^{-19}$ s.

However, depending on composition and dielectric properties, the characteristic time scale for streamer growth within electrically stressed oils ranges from the order of $10^{-10}$ s to the order of $10^{-6}$ s [30]. Evidently, the charge relaxation time for Gr in oil is at least 10 orders of magnitude smaller than the streamer development time scale for the oil. Consequently, Gr flakes tend to capture or scavenge the ionic entities released as a consequence of ionization of the oil much earlier than the charged entities can propagate and lead to the growth of the streamer (Fig. 8 (a)). As a result, the probability of DB at the expected value is drastically reduced, providing the oil with increased electrical stress bearing capacity.

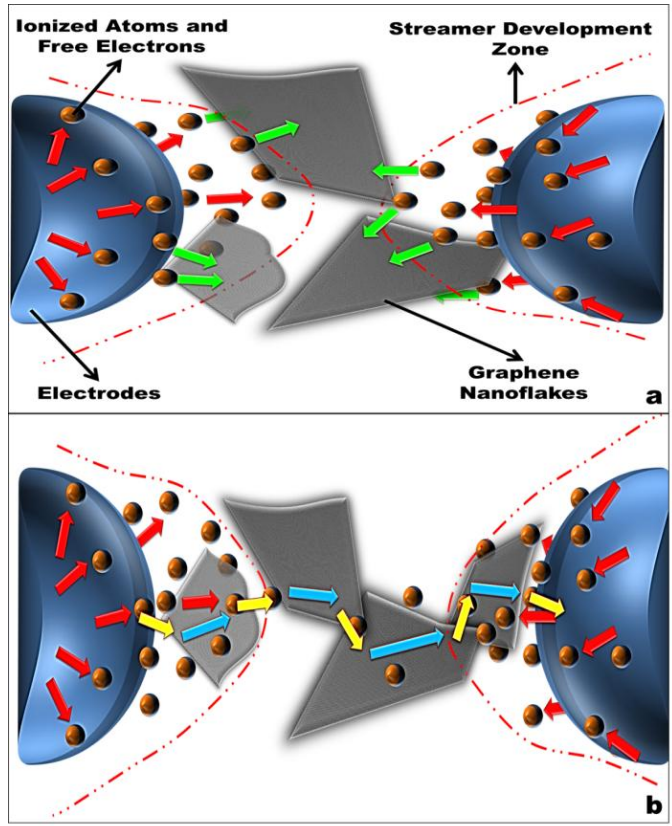

**Figure 8:** Qualitative and scale exaggerated illustration of the mechanism of scavenging charge released due to ionization of the oil molecules in the vicinity of the electrodes (red arrows indicate direction of traverse) **(a)** Dilute suspensions effectively capture the charges (indicated by green arrows) and delay the merging of the two streamers, thereby enhancing breakdown strength. **(b)** Above a critical concentration, the inter–flake distance between Gr is reduced to the point wherein the liberated charges can traverse through the flakes (indicated by blue arrows), 'hop' on the next flake (yellow arrows) due to the proximity and/or direct contact and reach the opposite streamer zone. The percolation chains thereby act as agents that induce 'shorting' of the steamers, often leading to numerous intermittent arc discharge and eventually leading to reduced breakdown strength.

Based on the fact that Gr/CNT has tens of orders of magnitudes larger electrical conductivity compared to insulating oils, Eqn. 1 can be restructured based on an infinite conductivity particle assumption. Since the relative permittivity of graphitic systems is very similar in magnitude to that of insulating oils and the electrical conductivity largely overshadows that of insulating oil, the restructured Eqn. 1 for the electric field at the nanostructure–oil interface at timescales much larger than charge relaxation time is expressible as

$$E_r(r = d_{np}/2, \theta) = 3E_0 \cos\theta \qquad (4)$$

As evident, the electric field intensity at the interface is thrice as strong as the incident field intensity within the system, leading to high concentration of converging field lines onto the nanostructures. As far as particles of lesser electrical conductivities are concerned, the interfacial field intensity is much weaker (compared to Eqn. 3) and therefore, presence of Gr/CNT leads to gross changes in the local electric field, leading to highly effective particle scavenging since the liberated electrons traverse along field lines, which in turn strongly converge onto such nanostructures. This explains the reason as to why even very minute concentrations of such nanostructures can lead to highly augmented insulation properties. For nanostructures with high dielectric constants, such as titanium dioxide and iron (II, III) oxide, the enhancement in breakdown voltage obtained is large, however, similar magnitudes as Gr/CNT systems is only realized at high concentrations (> 0.5 wt. %) [24, 27]. Since the electrical conductivities of such systems are magnitudes lower than GR/CNT, a much larger population of particles is required to achieve similar enhancements. However, these materials also possess high dielectric constants which lead to modification of the local permittivity of the nano–oil to aid the scavenging process, and thereby similar enhancements can be obtained at moderately high concentrations.

As the nanostructure population scavenges liberated electrons from the streamers, the average charge density per nanostructure approaches the saturation density. When the average charge density per nanostructure reaches the saturation density, the population is unable to scavenge further, and breakdown occurs. Consequently, with increasing population of nanostructures, i.e. with increased concentration, the amount of charge scavenged by the population before the average charge per nanostructure exceeds the saturation magnitude is higher than a low concentration case. As a result, increased concentration leads to enhanced magnitude of augmentation in breakdown voltage. However, although the enhancement is intuitively expected to be a linear function of growing concentration, in reality the enhancement follows a decaying growth rate, as evident from Figs. 5 and 7. Moreover, unlike titanium dioxide and iron (II, III) oxide based nano–oils [24, 27, 28] (although the exhibit enhancement is higher for Gr/CNT) the growth of enhancement saturates within dilute regimes. The charging dynamics of the nanostructure population can be considered analogous to the charging of a capacitive system, i.e. follows an exponentially decaying growth pattern. Ionization of fluid molecules leads to formation of both mobile free electrons as well as positively charged molecules. However, since the massive nature of the molecules lead to very low mobility compared to the electrons, the nanostructures come in contact with the electrons more frequently than positive ions, leading to predominant scavenging of electrons. Consequently, the growths of negative charge on the nanostructures grow with time. Gr and CNT, both nanomaterials with high free electron densities, thereby are soon overcrowded with electrons, resulting in a high negative charge density on the nanostructures. Consequently, they repel the incoming electrons which results in decaying charge scavenging efficacy in addition to the reduced potential of the nanostructure approaching its saturation point. With increasing population, the effective repulsion increases and thus the growth of enhancement decays out like a charging capacitor to a concentration where the charge acceptance capability of the system reaches saturation. Since the electron density in titanium oxide or iron oxide is much low compared to that of Gr/CNT, the resultant repulsion of inbound electrons are much smaller in magnitude at the same level of charging. As such, these systems can acquire charges up to greater concentrations and the growth of enhancement in such systems is often linear or bell shaped. However, if Gr/CNT systems cannot lead to enhanced breakdown beyond a certain threshold concentration, one would expect the DB voltage to attain a constant value for all concentrations beyond the threshold. However, sharp decrease in enhancement is observed and the same has been discussed in the next section.

Studies have also been performed to understand the nature in which the presence of nanostructures affects the insulating performance of the oil at elevated temperatures. Understanding such a phenomenon is of utmost importance technologically since most HV electrical devices generate large quantities of heat, thereby causing the cooling oil to perpetually operate at elevated temperatures compared to the ambient. Experiments reveal increases insulating performance of the base oil with increasing temperature, as illustrated in Fig. 9 and a thermal fluctuation based explanation can be conceived for the same. With increasing temperature, the viscosity of the oil reduces exponentially whereas the thermal energy of the liberated electrons and charged positive ions increases linearly. The enhanced thermal energy leads to enhanced amplitude of thermal fluctuations and thermal velocity experienced by the charged entities, resulting in disruption of the formative zones of the streamers. Owing to increased thermal randomness of the charges, the stability of the growth and development of the streamers is delayed, leading to delayed merger and augmented insulating capabilities. However, it has been observed (Fig. 9(a) and (b)) that addition of nanostructures leads to additional enhancement in breakdown voltage with increasing temperature, a phenomenon which can also be explained based on increased thermal energies of the nanostructures. As the temperature rises, the nanostructures experience enhanced degrees of Brownian randomness within the fluidic domain, leading to reduced interparticle collision mean free path and increased particle interactions. The increased interaction and randomness increases the degree of chaotic motion within the fluid, which supplemented with the increased instability of the charged entities, leads to further deterioration of the streamer stability. Consequently, the stable growth of the streamers is hampered and their merger delayed, leading to further augmented DB strength.

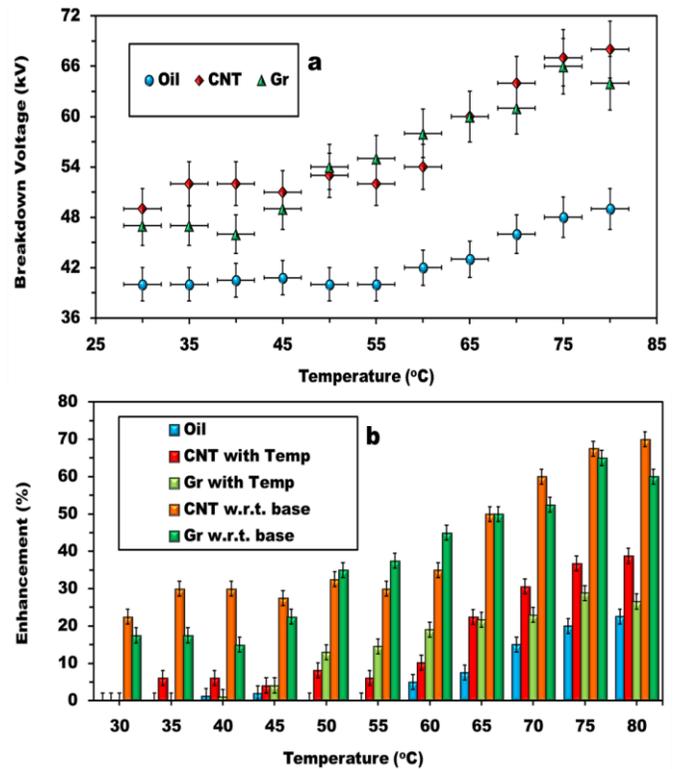

**Figure 9:** Dielectric breakdown characteristics of nano–oils with respect to temperature exhibits interesting phenomena governed by electro–thermal diffusion of nanostructures within the oil **(a)** Response of DB strength of Gr and CNT based nano–oils (0.01 wt. %) compared to that of the base oil as a function of temperature **(b)** Enhancement of DB strength of various nano–oils as a function of temperature, with respect to the base oil at a constant temperature (referred as 'with temp') as well as the base oil at ambient temperature (referred as 'w.r.t base').

### 3.2. Analytical Formulation

Although the mathematical analysis and associated discussions presented in the preceding section provides a comprehensive physics of the mechanisms by which the presence of highly conductive nanostructures such as Gr and CNT can lead to efficient augmentation of the DB voltage of liquids even at low nanostructure concentrations, when viewed at from an application or engineering point of view, an simplistic analytical formulation to derive the magnitude of augmentation is of utmost importance, since such a methodology would enable efficient and economic design of nano–oils for application fronts just based on theoretical formulation. The present section describes a dimensionally consistent and physically coherent analytical model which can be used at ease to evaluate the enhanced value of DB voltage for a given mineral oil or similar fluid in the event that the DB voltage of the base oil, size distribution and concentration of Gr and CNT structures, dielectric and electrical properties of the base oil and values of similar material properties are known. The model provides further insight into the mechanics of the phenomena from a mathematical point of view and has been derived based on simplifications of the field distribution equation pertaining to perfectly conducting nanostructures. The same is justified since the electrical conductivities of Gr and CNT at room temperature are approximately 20 orders of magnitude higher than that of the transformer oils and hence they behave similar to nanostructures with infinite conductivity when dispersed in such insulating media.

The anchoring philosophy behind determining the augmentation lies in comprehending the magnitude of streaming charged entities scavenged by the nanostructure population in the suspension. As the nano–oil is subjected to increasing electrical stresses, the oil molecules in the vicinity of the electrodes are ionized and the stripped off electrons lead to formation of streamers that try to propagate towards the opposite electrode. However, the Gr/CNT nanostructures in the vicinity of the electrodes begin scavenging the free electrons, thereby delaying the merging of the streamers and eventually breakdown. As the charge scavenging by the nanostructure initiates, the net charge on it grows with time, which in turn reduces its electron scavenging efficacy due to repulsive forces exerted by the scavenged electrons. The transient growth of charge ($Q(t)$) on the infinitely conducting nanostructure consequentially affects the localized electric field at the oil–nanomaterial interface and its radial and polar components can be mathematically expressed as [30]

$$E_r = E_0 \left[ \left(1 + \frac{d_{np}^3}{4r^3}\right) \cos\theta + \frac{Q(t)}{\pi \varepsilon_f d_{np}^2} \right] \quad (5)$$

$$E_\theta = E_0 \left( -1 + \frac{d_{np}^3}{8r^3} \right) \sin\theta \quad (6)$$

Scavenging of electrons by a given Gr/ CNT is possible only under the circumstance that there exists regions on the nanostructure surface wherein the net charge is predominantly positive in comparison to the incoming 'volley' of external electrons, such that the region of surface can absorb the inbound electrons without repulsion hampering the scavenging process. With increasing surface charge accumulation due to scavenging, the polar window available for absorbance of charges gradually diminishes. Mathematically, charge capture is possible at those regions of the nanostructure–oil interface ($r=d_{np}/2$) where the net electric field is finitely positive ($E_r \geq 0$). Evaluating Eqn. (4) in accordance to the described Dirichlet condition yields an expression for the polar window available ($\theta$) for electron scavenging at a given field intensity and nanostructure size as

$$\left| \frac{Q(t)}{3\pi \varepsilon_f d_{np}^2 E_0} \right| \leq \cos\theta \quad (7)$$

The delayed breakdown is promoted by the capture of the streamer forming charges by the nanostructures, thereby delaying the merger of the streamers beyond the usual breakdown voltage. The instant at which the nanostructure population in the vicinity of the streamer development regions is saturated and unable to scavenge further, the streamers develop au natural and leads to DB. Determining the enhancement requires an estimate of the scavenging capacity of the nanostructures utilized. At the point of saturation for a nanostructure, the interface is unable to capture any more external charge, i.e. mathematically, the polar window available for charge scavenging decays down to zero ($\theta=0$). It is at this instant that the charge gathered by a single nanostructure can be termed as the saturation charge ($Q_{S,np}$) and can be accordingly deduced from Eqn. (6) as

$$Q_{S,np} = 3\pi \varepsilon_f d_{np}^2 E_0 \quad (8)$$

For mono–disperse systems involving spherical nanostructures with the diameters confined to a tight Gaussian distribution, the usage of a simple mean diameter suffices for the present endeavor. However, for highly poly–disperse and high aspect ratio such as Gr and CNT respectively, the effective size characterizing the system requires being determined. For a low or near unity aspect ratio but high poly–dispersity index system such as Gr (assuming the flakes to be near squares), the characteristic size is obtained from the DLS data (Fig. 1(c)). Based on the approximate intensity of a particular peak ($I$), the flake size corresponding to the peak position ($d$) and the number of peaks ($m$), a weighted mean representative flake size ($d_{eq}$) can be deduced as

$$d_{eq,Gr} = \frac{\sum_{j=1}^{m}(d_j I_j)}{\sum_{j=1}^{m} I_j} \quad (9)$$

Since the Gr based system is a flake based population, the area exposed to electron scavenging is taken care of once the representative flake size has been determined. However, for a low poly–dispersity index and high aspect ratio (AR) system such as CNT (low poly–dispersity with reference to the outer diameter), the representative diameter can be effectively modeled based on equivalence of the surface area of a nanotube to that of a hypothetical spherical particle. The surface area based equivalence is of prime importance since it is the only method to ensure uniform scavenging by the nanotubes for the analysis, which otherwise scavenge in accordance to its orientation with respect to the streamers. The equivalent diameter for a CNT population can be thus determined as

$$d_{eq,CNT} = 2d_{CNT}\sqrt{AR} \qquad (10)$$

Furthermore, the expression in Eqn. (7) is devoid of the dielectric property of the nanostructure and unless modified to cater to the same, provides accurate predictions only for cases wherein the relative permittivities of the fluid and the nanomaterial are of similar magnitudes. However, for situations not pertaining to the above condition (which is expected to be a majority since solids often possess higher dielectric constants than oils), a simplistic formulation known as Matijevic equation [34] expressed in Eqn. (10), can be utilized to obtain the effective relative permittivity of the nano–oil from the relative permittivities of its constituents and nanomaterial concentration ($\varphi$, wt./wt.), and the same maybe utilized in lieu of the dielectric constant of the fluid phase in Eqn. (7) to obtain a modified expression for saturation charge.

$$\varepsilon_{eff} = \varepsilon_f + \frac{3\varepsilon_f(\varepsilon_{np} - \varepsilon_f)}{2\varepsilon_f + \varepsilon_{np}}\varphi \qquad (11)$$

In order to estimate the number density of the net charge scavenged the spatial distribution of Gr/ CNT in the oil should be determined. Assuming a uniform distribution of nanostructures in the oil phase, the number density of nanostructures ($n_{np}$) can be evaluated from the nanomaterial concentration, densities ($\rho$) of the fluid and nanomaterial phases and approximate volume of a single representative nanostructure ($v_{snp}$) as

$$n_{np} = \frac{\rho_f \varphi}{\rho_{np} v_{snp}} \qquad (12)$$

Moreover, scavenging of free electrons is not the sole mechanism by which the DB strength is enhanced in such systems. As the electrons are stripped off the oil molecules, the molecules attain positive charges, which lead to formation of independent positive streamers and can also lead to DB [33]. As the nanostructures scavenge electrons, the net negative charge on them grows and their propensity of attraction towards the positive streamer grows, leading to disruption of formation of the positive streamers, which further reduces the possibility of DB. Also, random thermal fluctuations of the nanostructures within the fluid medium enhances the window of scavenging since a larger traversal region signifies exposure to a greater number of free electrons. Consequently, the ability of the nanostructure to diffuse across the fluid space due to the acquired electro–thermal fluctuations need to be incorporated within the formulation. Considering a single Gr flake or CNT, the effective electro–thermal mobility ($\mu_{e-t}$) is expressible as a function of the acquired charge ($q$), intrinsic thermal energy (in terms of the Boltzmann constant $k_B$ and absolute temperature $T$) and Brownian diffusivity ($D_B$) as [34]

$$\mu_{e-t} = \frac{qD_B}{k_B T} \qquad (13)$$

For quantification of the mobility, an estimate of the charge scavenged by each particle is necessary. Since the present formulation is analytical in nature, the charging dynamics is irrelevant as the final outcome desired is to determine the degree of enhancement, a time by which the nanostructure population has acquired saturation charge. Thereby, the magnitude of saturation charge can be utilized in the present formulation. Furthermore, since the characteristic sizes are in the micro–nanoscales, the nanostructures can be assumed to behave similar to spherical particles whose hydrodynamics in the fluid media can be represented accurately by Stokesian dynamics. Thereby, the Stokes–Einstein formulation for Brownian diffusivity [18] can be utilized to obtain the expression for the electro–thermal mobility of a representative nanostructure as

$$\mu_{e-t} = \frac{|Q_{s,np}|}{3\pi\eta d_{np}} \qquad (14)$$

In the above equation, the diameter of nanoparticle should be replaced by the equivalent diameter for Gr and CNT, as obtained from earlier discussions. As evident, the viscosity of the fluid decreases with increasing temperature, thereby leading to augmented mobility of the nanostructures, which in turn leads to enhanced probability of scavenging, leading to enhanced DB strength at elevated temperatures (Fig. 8). However, the efficiency of the diffusive form of hampering positive streamers can be as high as the magnitude by which the nanoparticle mobility is greater than the positive charge mobility ($\mu_C$) (in general of the order of $1\times10^{-9}$ m$^2$V$^{-1}$s$^{-1}$). Thereby, the ratio of the two diffusivities governs the mechanism. Finally, the maximum possible number density of charge which can be scavenged ($\chi_{np}$) from the streamers by the involved nanostructure population before attaining the state of saturation is expressible as

$$\chi_{np} = n_{np} Q_{S,np}\left(\frac{\mu_{e-t}}{\mu_c}\right) \qquad (15)$$

In insulating fluids such as oils, the mechanism of breakdown is governed by the ionization and subsequent release of free electrons which leads to the formation of a conducting ambience beyond a certain intensity of electrical stressing. Thus, the DB strength of any oil can be expected to hold an inverse relationship with its maximum free charge density, i.e. the magnitude of charge per unit volume which can be stripped off from its bound form when subjected to electrical stresses and it can be intuitively argued that breakdown of the matter occurs when the free charge density attains a fraction of its maximum limit; the fraction being a constant for a particular material. The nanomaterials present in the nano–oils effectively scavenge and absorbs a portion of the charge population required to bring about DB, thereby delaying the process. The enhancement achieved thereby depends on the magnitude of charge scavenged per unit volume by the nanomaterials up to the point of electrical stressing where the base oil is expected to yield and undergo breakdown. Substituting the electric field intensity (the ratio of the breakdown voltage to the electrode spacing) at which the base oil undergoes DB ($V_f$) in Eqn. (14), the saturation charge density of the nanostructure population at that level of electrical stress is determinable. Since a portion of the required charge density is scavenged, breakdown does not occur at the DB voltage of the base oil. Therefore, enhanced electrical stress is required to ionize the oil further so as to cause DB. Based on the maximum free charge density of the fluid ($\chi_f$) (~ 550 Cm$^{-3}$ and 450 Cm$^{-3}$ for Oil1 and Oil2 respectively, as determined from linear extrapolation from known data [30]), the probability ($p(b)$) by which breakdown at the expected point has been delayed can be expressed as

$$p(b) = \frac{\chi_{np}}{\chi_f} \quad (16)$$

Thereby, an estimate of the probability by which the oil does not undergo DB at the expected voltage due to scavenging by the dispersed Gr/CNT can be made from Eqn. (15). Hence, it may be argued that the probability by which the breakdown is averted at the expected field intensity is the probability by which the magnitude of DB voltage is enhanced over and above the same. Although DB is a complex non–linear dynamical phenomenon; the linearized argument put forward can be utilized for ease of mathematical formulation without suffering appreciable deviation, as observed later from the validations. In accordance to the mechanism proposed, the DB voltage of the nano–oil ($V_n$) is expressible as

$$V_n = V_f(1 + p(b)) \quad (17)$$

Based on the proposed mathematical formulation, the predicted DB voltages for the various samples of nano–oils have been validated against experimental findings and accurate predictability has been observed. The validation plots for Gr and CNT based nano–oils have been illustrated in Fig. 10(a) and (b) respectively. It is observed that the prediction curves are nearly linear functions of nanostructure concentration; however, experimental data holds the quasi–linear growth up to certain concentration and the enhancement decays beyond that. In reality, the decreasing stability of the suspension in conjunction with the propensity of hastened corona discharge between the electrodes caused by percolation chain structures at higher nanostructure concentrations leads to decrement of the magnitude of enhancement of the DB voltage. With increasing concentration, the problems associated with the stability of the suspensions creep in and minimal electrodynamic stressing can often lead to hastened flocculation, thereby leading to inefficient enhancement of DB voltage. Furthermore, another mechanism initiates hampering the augmentation. At higher concentrations, the flake/tube density is high and thereby the separation between two nanostructures decreases drastically.

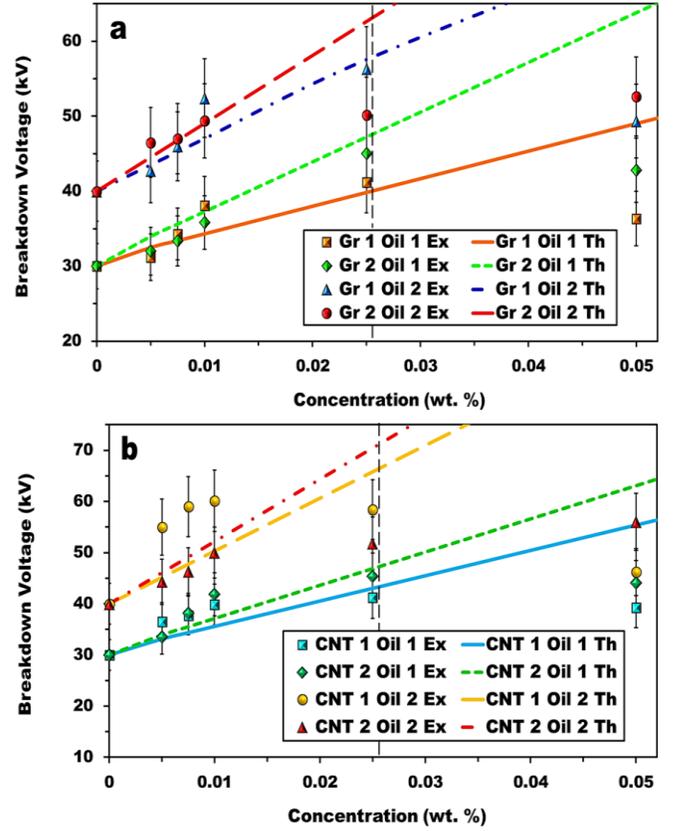

**Figure 10:** Validation of the predictions obtained from the theoretical model against experimental data for **(a)** Gr based nano–oils **(b)** CNT based nano–oils. Owing to stability issues of the colloidal phase beyond certain critical concentrations as well as additional breakdown favoring phenomena such as percolation chains creeping in, the predictability of the model is restricted to the ascending regions of the augmented voltages. The dotted vertical lines provide an estimate of the critical concentrations, beyond which the model fails to capture the phenomena, neither qualitatively nor quantitatively.

Since the electronic mobility within Gr/CNT is very high, the small separation helps in formation of a conducting chain within the streamer development zones. These chains are theorized to be formed by the closely arranged nanostructures and linked by the free electrons within the streamer zones. At low concentrations, the inter–nanostructure spacing is too large for the free electron population in the vicinity to link them electrically. Qualitative evidence of such a phenomenon has been observed, such that, the number of intermittent discharge between the two electrodes, even at low electric field intensities, is high for higher concentrations, leading to hastened breakdown. The reason behind such large number of weak, intermittent discharges before final breakdown can be attributed to the percolation chain theory. Thereby; due to the dual phenomena of decreased stability and '*short–circuit*' initiating chained structures, only a finite degree of enhancement of the dielectric strength is possible. Further, the probability of percolating networks formed by Gr flakes in direct contact increases the probability of the 'shorting' behavior. An increased contact based percolation network is possible at higher concentrations of flakes/tubes exactly above a certain population density. As evident from Figs. 5 and 7 the enhancement in breakdown strength suffers a massive drop immediately beyond the critical concentrations, thereby adding credibility to the proposed hypothesis. As the analytical model does not encompass all such mechanisms, it continually predicts increasing DB voltage as a function of concentration. However, it accurately predicts the voltage up to the critical concentration (indicated by vertical dotted lines in Fig. 9), i.e. the point beyond which the enhancement decays, and can therefore be used as an efficient tool to design Gr/CNT based nano–oils for high charge density electrical machines and drives.

## 4. Conclusions

To infer, the present article reports for the very first time, the utility of Gr nanoflakes and CNT as agents that render highly augmented dielectric breakdown strengths to insulating fluids, such as transformer oils. Gr and CNT based dilute nano–oils, synthesized by dispersing minimal quantities of the nanostructures in graded transformer oils have been synthesized and their breakdown characteristics experimentally studied as per standard codes and protocols, utilizing a high voltage fluid testing device. High degrees of enhancement in the DB voltage have been obtained for all the tested categories of nano–oils, thereby confirming the claims put forward. A mathematical analysis based on the electro–hydrodynamics of nanostructures in fluidic phases has been utilized to explain the mechanisms and physics behind the process. A detailed and comprehensive explanation for the various phenomena observed as functions of system and material properties has been presented. The mathematical analysis has also been evolved into an analytical formulation which can accurately predict the breakdown strengths of such nano–oils when the various system properties are known inputs. The present work thus concretely establishes the potential of Gr and CNTs in elevating the operating capabilities and mitigating damage and safety concerns of high voltage electrical systems cooled by insulating fluids.

## Acknowledgments

The authors thank the Defense Research and Development Organization (DRDO) for funding the present work. They also thank the Sophisticated Analytical Instruments Facility (SAIF), IIT Madras and the Central Electronics Centre (CEC), IIT Madras for material characterizations. PD also thanks the Ministry of Human Resource and Development (MHRD), Govt. of India, for the doctoral research scholarship.

## References

[1] T. Mueller, F. Xia, and P. Avouris, "Graphene photodetectors for high-speed optical communications," *Nature Photonics,* vol. 4, pp. 297 - 301, 2010.

[2] Y. G. Semenov, J. M. Zavada, and K. W. Kim, "Electrically controlled magnetic switching based on graphene-magnet composite structures " *Journal of Applied Physics,* vol. 107, pp. 064507, 2010.

[3] H. Pu, and F. Jiang, "Towards high sedimentation stability: magnetorheological fluids based on CNT/Fe3O4 nanocomposites " *Nanotechnology,* vol. 16 pp. 1486, 2005.

[4] G. Lu, L. E. Ocola, and J. Chen, "Reduced graphene oxide for room-temperature gas sensors " *Nanotechnology,* vol. 20 no. 445502, pp. 445502, 2009

[5] C. C. Su, T. Liu, N. K. Chang, B. R. Wang, and S. H. Chang, "Two dimensional carbon nanotube based strain sensor," *Sensors and Actuators A: Physical,* vol. 176, pp. 124-129, 2012.

[6] M. Zhou, Y.-H. Lu, Y.-Q. Cai, C. Zhang, and Y.-P. Feng, "Adsorption of gas molecules on transition metal embedded graphene: a search for high-performance graphene-based catalysts and gas sensors " *Nanotechnology,* vol. 22, no. 385502, pp. 385502, 2011

[7] S. M. Kim, E. B. Song, S. Lee, S. Seo, D. H. Seo, Y. Hwang, R. Candler, and K. L. Wang, "Suspended few-layer graphene beam electromechanical switch with abrupt on-off characteristics and minimal leakage current " *Applied Physics Letters,* vol. 99, pp. 023103, 2011.

[8] S.-E. Zhu, M. K. Ghatkesar, C. Zhang, and G. C. A. M. Janssen, "Graphene based piezoresistive pressure sensor " *Applied Physics Letters,* vol. 102, pp. 161904, 2013.

[9] F. Schwierz, "Electronics: Industry-compatible graphene transistors," *Nature,* vol. 472, pp. 41-42, 2011.

[10] S. Rumyantsev, G. Liu, W. Stillman, M. Shur, and A. A. Balandin, "Electrical and noise characteristics of graphene field-effect transistors: ambient effects, noise sources and physical mechanisms," *Journal of Physics: Condensed Matter,* vol. 22 pp. 395302, 2010.

[11] J. Y.Q., Z. Q., and L. L., "Planar MEMS Supercapacitor using Carbon Nanotube Forests," *IEEE 22nd International Conference on Micro Electro Mechanical Systems, 2009. MEMS 2009. ,* pp. 587-590 2009.

[12] S.-W. Kim, D.-H. Seo, H. Gwon, J. Kim, and K. Kang, "Fabrication of FeF3 Nanoflowers on CNT Branches and Their Application to High Power Lithium Rechargeable Batteries," *Advanced Materials,* vol. 22(46), pp. 5260–5264 2010.

[13] E. Antolini, "Graphene as a new carbon support for low-temperature fuel cell catalysts," *Applied Catalysis B: Environmental,* vol. 123–124, pp. 52–68, 2012.

[14] Z. Ming, A. Jagota, E. D. Semke, B. A. Diner, R. S. McLean, S. R. Lustig, R. E. Richardson, and N. G. Tassi, "DNA-assisted dispersion and separation of carbon nanotubes," *Nature Materials,* vol. 2(5), pp. 338-342, 2003.

[15] A. Schierz, and H. Zänker, "Aqueous suspensions of carbon nanotubes: Surface oxidation, colloidal stability and uranium sorption," *Environmental Pollution,* vol. 157, no. 4, pp. 1088-1094, 4//, 2009.

[16] Y. Chen, L. Chen, H. Bai, and L. Li, "Graphene oxide-chitosan composite hydrogels as broad-spectrum adsorbents for water purification," *Journal of Materials Chemistry A,* vol. 1, no. 6, pp. 1992-2001, 2013.

[17] P. Dhar, S. Bhattacharya, S. Nayar, and S. K. Das, "Anomalously augmented charge transport capabilities of biomimetically transformed collagen intercalated nanographene-based biocolloids," *Langmuir,* vol. 31, no. 12, pp. 3696-706, Mar 31, 2015.

[18] P. Dhar, S. S. Gupta, S. Chakraborty, A. Pattamatta, and S. K. Das, "The role of percolation and sheet dynamics during heat conduction in poly-dispersed graphene nanofluids," *Applied Physics Letters,* vol. 102, no. 16, pp. 163114, 2013.

[19] P. Dhar, M. H. D. Ansari, S. S. Gupta, V. M. Siva, T. Pradeep, A. Pattamatta, and S. K. Das, "Percolation network dynamicity and sheet dynamics governed viscous behavior of polydispersed graphene nanosheet suspensions," *Journal of nanoparticle research,* vol. 15, no. 12, pp. 1-12, 2013.

[20] N. V. Sastry, A. Bhunia, T. Sundararajan, and S. K. Das, "Predicting the effective thermal conductivity of carbon nanotube based nanofluids," *Nanotechnology,* vol. 19, no. 5, pp. 055704, 2008.

[21] M. Chiesa, and S. K. Das, "Experimental investigation of the dielectric and cooling performance of colloidal suspensions in insulating media," *Colloids and Surfaces A: Physicochem. Eng. Aspects* vol. 335 pp. 88–97, 2009.

[22] J. G. Hwang, M. Zahn, Francis M. O'Sullivan, L. A. A. Pettersson, O. Hjortstam, and R. Liu, "Electron Scavenging by Conductive Nanoparticles in Oil Insulated Power Transformers," *2009 Electrostatics Joint Conference,* vol. 1.1, pp. 1-12, 2009.

[23] D. Yue-fan, L. Yu-zhen, W. Fo-chi, L. Xiao-xin, and L. Cheng-rong, "Effect of TiO2 nanoparticles on the breakdown strength of transformer oil," *Conference Record of the 2010 IEEE International Symposium on Electrical Insulation (ISEI),,* vol. ISEI 2010, pp. 1-3, 2010.

[24] Y. Du, Y. Lv, C. Li, M. Chen, Y. Zhong, J. Zhou, X. Li, and Y. Zhou, "Effect of Semiconductive Nanoparticles on Insulating Performances of Transformer Oil," *IEEE Transactions on Dielectrics and Electrical Insulation,* vol. 19(3), pp. 770-776, 2012.

[25] Z. Jian-quan, D. Yue-fan, C. Mu-tian, L. Cheng-rong, L. Xiao-xin, and L. Yu-zhen, "AC and Lightning Breakdown Strength of Transformer Oil Modified by Semiconducting Nanoparticles," *2011 Annual Report Conference on Electrical Insulation and Dielectric Phenomena (CEIDP),* vol. CEIDP 2011, pp. 652 - 654 2011.

[26] D.-E. A. Mansour, E. G. Atiya, R. M. Khattab, and A. M. Azmy, "Effect of Titania Nanoparticles on the Dielectric Properties of Transformer Oil-Based Nanofluids," *2012 Annual Report Conference on Electrical Insulation and Dielectric Phenomena (CEIDP),* vol. CEDIP 2012, pp. 295-298, 2012.

[27] J.-C. Lee, H.-S. Seo, and Y.-J. Kim, "The increased dielectric breakdown voltage of transformer oil-based nanofluids by an external magnetic field," *International Journal of Thermal Sciences,* vol. 62, pp. 29-33, 2012.

[28] V. Segal, A. Rabinovich, D. Nattrass, K. Raj, and A. Nunes, "Experimental study of magnetic colloidal fluids behavior in power transformers," *Journal of Magnetism and Magnetic Materials* vol. 215-216, pp. 513-515, 2000.

[29] J. Li, Z. Zhang, P. Zou, S. Grzybowski, and M. Zahn, "Preparation of a Vegetable Oil-Based Nanofluid and Investigation of Its Breakdown and Dielectric Properties," *IEEE Electrical Insulation Magazine,* vol. 28(5), pp. 43-50, 2012.

[30] J. G. Hwang, M. Zahn, F. M. O'Sullivan, L. A. A. Pettersson, O. Hjortstam, and R. Liu, "Effects of nanoparticle charging on streamer development in transformer oil-based nanofluids," *Journal of Applied Physics,* vol. 107, no. 014310, pp. 014310, 2010.

[31] N. Kurra, V. Bhadram, C. Narayana, and G. Kulkarni, "Few layer graphene to graphitic films: infrared photoconductive versus bolometric response," *Nanoscale,* vol. 5, pp. 381–389, 2013.

[32] P. K. Watson, W. G. Chadband, and M. Sadeghzadeh-Araghi, "The role of electrostatic and hydrodynamic forces in the negative-point breakdown of liquid dielectrics," *IEEE Transactions on Electrical Insulation,* vol. 26(4), pp. 543-550, 1991.

[33] A. Beroual, M. Zahn, A. Badent, K. Kist, A. J. Schwabe, H. Yamashita, K. Kamazawa, M. Danikas, W. G. Chadbrand, and Y. Torshin, "Propagation and Structure of Streamers in liquid Dielectrics," *IEEE Electrical Insulation Magazine,* vol. 14(2), pp. 6-17, 1998.

[34] P. Dhar, A. Pattamatta, and S. K. Das, "Trimodal charge transport in polar liquid-based dilute nanoparticulate colloidal dispersions," *Journal of Nanoparticle Research,* vol. 16, no. 10, pp. 1-20, 2014.

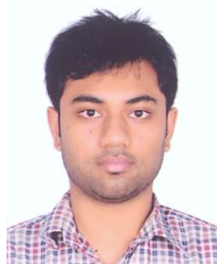

**Purbarun Dhar** was born in Calcutta, India in May, 1990. He received a B. Tech in Mechanical Engineering in 2012 from the National Institute of Technology at Durgapur, India. He submitted his thesis for M.S.–Ph.D. dual degree to the Department of Mechanical Engineering of the Indian Institute of Technology at Madras, India, in September 2015. Since then, he is associated with the institute as an Institute Pre-doctoral Fellow. His doctoral work is focused upon developing novel colloidal systems involving nanomaterials, mostly graphene and CNT, and probing the various augmented thermophysical and electromagnetic transport properties of such colloidal systems. His other research interest lies in biomedical nanotechnology for therapeutics, micro-nanoscale transport processes, engineering multiphysics and biomimetic systems and has published research articles in all the fields of interest in journals of AIP, RSC, ACS, IEEE, Springer Verlag etc.

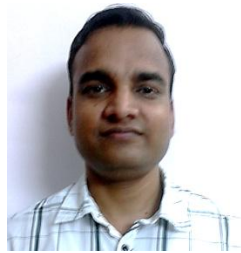

**Ajay Katiyar** was born in Farrukhabaad, India in July 1976. He received a B. Engg in Mechanical Engineering from the Pt. Ravi Shankar Shukla University, Raipur, India in 2001 and M.Tech in Manufacturing Engineering from the UP Technical University, Lucknow, India in 2005. He was employed as an engineer from Bharatpur Agriculture Implement Works, Bharatpur, India from 2001 to 2003. He was a Senior Lecturer, at the Maharana Pratap Engineering College, Kanpur, India from 2005 to 2007 and Senior Lecturer, CSJM University Kanpur, India from 2007 to 2008. Since May 2008, he is a Scientist with the Defence Research and Development Organization (DRDO), India. Since 2014, he is also pursuing a Ph.D. at the Indian Institute of Technology Madras, India. His areas of research interest are magneto and electrorheology in nanofluids, nano-finishing, industrial and automotive applications of nanofluids etc.

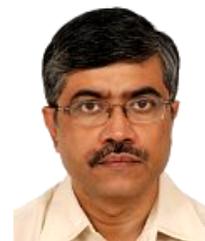

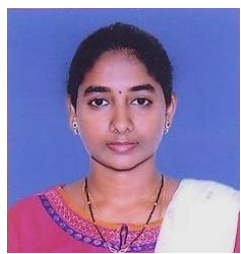

**Lakshmi Sirisha Maganti** was born in Sattenapalli, India in August 1987. She received a B.Tech in Mechanical Engineering from Nagarjuna University, Guntur, India and a M. Engg (with University Gold Medal) in Refrigeration and Air Conditioning from the College of Engineering, Anna University, Chennai, India. She was employed as an Assistant Professor of Mechanical Engineering at the SKR Engineering College, Chennai, India from 2012 to 2013. Since 2013, she is pursuing a Ph.D. at the Department of Mechanical Engineering of the Indian Institute of Technology Madras, India. Her research interests include Computational Fluid Dynamics and Heat Transfer, nanofluids, microfluidic systems, etc.

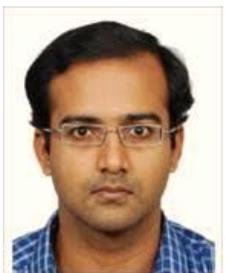

**Arvind Pattamatta** is as an Assistant professor in the Department of Mechanical Engineering at Indian Institute of Technology Madras since 2010. He received his Bachelor, Master's and Doctoral degrees in Aerospace Engineering from the University of Madras in 2001, the Indian Institute of Science in 2003, and the State University of New York at Buffalo in 2009, respectively. From 2003 till 2005, he was employed as a Design Engineer in the Combustion Center of Excellence at GE India Technology Center in Bangalore, India where he was using Computational Fluid Dynamics based tools for the analysis of fluid flow and heat transfer in GE Aircraft engine Diffusers and Combustion chambers. During the years 2009-2010 he was a Principal Scientist in the R.D. Aga Research, Technology and Innovation Centre at Thermax Limited, Pune. He is the recipient of Alexander von Humboldt fellowship for the year 2013 to conduct research studies at TU Darmstadt, Germany. His research interests are in the areas of Computational Nanoscale energy Transport, Computational Fluid dynamics, microfluidics, Turbulence Modeling, and High performance computing.

**Sarit Kumar Das** is a Professor with the Mechanical Engineering Department of Indian Institute of Technology Madras, India. Since July 2015, he is also a Professor at the School of Mechanical, Materials and Energy Engineering (SMMEE) and the Director of Indian Institute of Technology Ropar, India. He has published four books and more than 200 research papers. His current research interests include heat transfer in nanofluids, microfluidics, biological heat transfer, nanoparticle mediated drug delivery in cancer cells, heat exchangers, boiling in mini/microchannels, fuel cells, jet instabilities, heat transfer in porous media, and computational fluid dynamics. He is a recipient of the DAAD and Alexander von Humboldt Fellowship of Germany. He is a fellow of the Indian National Academy of Engineering and the National Academy of Sciences, India. He has been awarded the Peabody Visiting Professorship at the Mechanical Engineering Department, Massachusetts Institute of Technology, Cambridge, USA, in 2011. He is the Editor-in-Chief of the International Journal of Micro/Nanoscale Transport and an Associate Editor of the journal Heat Transfer Engineering.